\documentclass
[twocolumn,showpacs,preprintnumbers,amsmath,amssymb,prc]{revtex4}
\usepackage{graphicx}
\usepackage{dcolumn}
\usepackage{bm}

\begin{document}

\title{Temperature of projectile like fragments in heavy ion collisions}

\author{S. Das Gupta$^1$, S. Mallik$^2$ and G. Chaudhuri$^{2}$}

\affiliation{$^1$Physics Department, McGill University, Montr{\'e}al, Canada H3A 2T8}
\affiliation{$^2$Theoretical Physics Division, Variable Energy Cyclotron Centre, 1/AF Bidhan Nagar, Kolkata 700064, India}

\date{\today}

\begin{abstract}
A model in which a projectile like fragment can be simply regarded as a remnant after removal of some part of the projectile leads to an excited fragment. This excitation energy can be calculated with a Hamiltonian that gives correct nuclear matter binding, compressibility and density distribution in finite nuclei. In heavy ion collisions the model produces a dependence of excitation energy on impact parameter which appears to be correct but the magnitude of the excitation energy falls short. It is argued that dynamic effects left out in the model will increase this magnitude. The model can be directly extended to include dynamics but at the expense of increased computation. For many calculations for observables, a temperature is an easier tool to use rather than an excitation energy. Hence temperature dependences on impact parameter in heavy ion collisions are displayed.
\end{abstract}


\maketitle
{\bf {\it Introduction:-}}
In recent times we proposed a model \cite{Mallik2, Mallik3, Mallik101} for projectile fragmentation whose predictions were compared with many experimental data with good success. The model has three parts. To start
with we need an abrasion cross-section.  This was calculated using straight line trajectories for the projectile and the target leading to a definite mass and shape to the projectile like fragment (PLF). The PLF will have an excitation energy.  It was conjectured that this will depend upon the relative size of the PLF with respect to the projectile, i.e., on $(A_s/A_0)$ where $A_s$ is the size of the PLF and $A_0$ is the size of the whole projectile.  Since the size $A_s$ of the PLF depends upon the impact parameter of the collision, dependence on $(A_s/A_0)$ means the excitation energy depends upon the impact parameter \cite{Mallik3, Mallik101}.  In our original version this dependence was neglected \cite{Mallik2} but became an important feature in the improved model \cite{Mallik3}.  Instead of excitation energy we use temperature $T$.  The hot PLF will disintegrate into different composites
which can be calculated using a canonical thermodynamic model (CTM) \cite{Das}. Evaporation from hot composites which result from CTM was implemented
\cite{Mallik1}.

In our model the temperature of the PLF was not calculated, it was fitted from data. In this Letter we try to estimate $T$ from a more basic approach.  We
are not trying here to formulate a complete model for projectile fragmentation. The earlier papers \cite{Mallik2,Mallik3,Mallik101} had that goal.  There
are models which calculate observables which can be directly compared with experiment.  One example is the heavy ion phase space exploration (HIPSE)
\cite{Lacroix} model.  Another model that has been used is the antisymmetrised molecular dynamics (AMD) model \cite{Ono}.  These are relevant to our
earlier papers rather than to the present work.

The concept of temperature has proved to be useful to
understand many features of projectile fragmentation.
Temperature $T$ in the PLF has been studied a great deal in the past using a
combination of theory and experimental data. A very popular method uses
experimental populations of excited states (for example, the
"Albergo" formula \cite{Albergo}) to deduce a temperature. Many data suggest
that the temperature is of the order of 5 MeV. There is also unmistakable
evidence that the temperature falls off with increasing impact parameter.
It was shown in \cite{Mallik3} that for beam energies between 140 MeV/n
and 1 GeV/n (the only cases that were tried)  a remarkably simple
parametrisation $T(b)=7.5MeV-[A_s(b)/A_0]4.5MeV$ worked well
for all the pair of ions.
In this Letter we are trying to see if such simple feature can be understood
in a transparent physical picture.  Simple models for PLF excitations have
also been made in
the past \cite{Gaimard,Brohm}.  The relationship of that work to ours will be
discussed in the last part of this Letter.

If straight line geometry is used then it is obvious that the PLF is created
with a crooked shape.  If the excitation energy in the PLF
is mostly due to its crooked shape at the time of separation, one can estimate
the excitation energy assuming a liquid drop model with a volume term, a
surface tension term and Coulomb
contribution.  One is probably confined to assuming a constant density which
is not realistic.

The objective of this Letter is to estimate the temperature in the PLF using a
more microscopic approach. The lowest order approximation we make to estimate
the temperature arises from the crooked shape that results from straight line
geometry for cuts.  The excitation energy originates from structure effects
solely as in a liquid drop
model.  But we are not using a liquid drop model nor are we limited to constant density. Our method can be extended to consider dynamic effects using a
transport model, specifically the BUU (Boltzmann-Uehling-Uhlenbeck) model.
The techniques we use, even for
the lowest order estimates, are well-known in BUU calculations.

We should hasten to add that the calculations reported in this Letter are not transport model calculations.  These are static calculations. There is no time evolution. The beam energy does not enter but for the validity of the model the beam should be sufficiently energetic  so that straight line trajectories are a valid approximation.  There are transport model calculations which compute excitation energy per nucleon for quasi-projectile and quasi-target \cite{Galichet} at 52 MeV per nucleon and 72 MeV per nucleon.  Our model would not apply at such low energy.  Our model here  was developed for cases like Sn on Sn at 600 MeV per nucleon, Xe on Pb at 1 GeV per nucleon.  The lowest energy we had in mind was 140 MeV per nucleon with Ni as the projectile.

We do calculation for Sn on Sn, Xe on Pb, Ni on Be and Ta as we have used our model for these cases before \cite{Mallik2, Mallik3, Mallik101} and experimental results are known \cite{Ogul, Henzlova, Mocko}.  In heavy ion collisions we consider one ground state nucleus hitting another ground state nucleus.  We first turn to the model of the ground state.

{\bf {\it Model of the ground state:-}}
We use  Thomas-Fermi (TF) solutions for ground states. Complete details of our procedure for TF solutions plus the choice of the interactions are given in Ref. \cite{Lee}. For completeness the prescription is outlined. The kinetic energy density is given by
\begin{equation}
T(\vec r)=\int d^3p f(\vec{r},\vec{p})p^2/2m
\end{equation}
where $f(\vec{r},\vec{p})$ is the phase space density.
Since we are looking for lowest energy we take, at each $\vec{r}$, $f(\vec {r},\vec{p})$ to be non-zero from 0 to some maximum $p_F(\vec{r})$.  Thus we will have
\begin{equation}
f(r,p)=\frac{4}{h^3}\theta [p_F(r,p)-p]
\end{equation}

The factor 4 is due to spin-isospin degeneracy and using the spherical symmetry of the TF solution we have dropped the vector sign on $r$ and $p$. This leads to
\begin{equation}
T=\frac{3h^2}{10m}[\frac{3}{16\pi}]^{2/3}\int \rho(r)^{5/3}d^3r
\end{equation}
For potential energy we take
\begin{eqnarray}
V=A\int d^3r\frac{\rho^2(r)}{2}+\frac{1}{\sigma+1}B\int \rho^{\sigma+1}({r})d^3r\nonumber\\
+\frac{1}{2}\int d^3rd^3r' v(\vec{r},\vec{r}')\rho(\vec{r})\rho(\vec{r'})
\end{eqnarray}

The first two terms on the right hand side of the above equation are zero range Skyrme interactions.  The third which is a finite range term is often suppressed and the constants $A,B,\sigma$ are chosen to fit nuclear matter equilibrium density, binding energy per nucleon and compressibility.  In heavy ion collisions, for most purposes, this will be adequate but for what we seek here, possibly a small excitation energy, this is wholly inadequate.  Thomas-Fermi solution is obtained by minimising $T+V$.  With only zero range force, $\rho(r)$ can be taken to be a constant which goes  abruptly to zero at some $r_0$ fixed by the total number of nucleons.  Now if $\rho$ is chosen to minimise the energy then, a nucleus, at this density with a cubic shape is as good as a spherical nucleus.  Besides the minimum energy nucleus will have a sharp edge, not a realistic density distribution.  This problem does not arise in quantum mechanical treatment with Skyrme interaction.  Including a finite range potential in TF one recovers a more realistic density distribution for the ground state and one regains the nuclear structure effects which will contribute to excitation the PLF.  This is discussed in more detail in Ref. \cite{Lee}.

We note in passing that Lenk and Pandharipande introduced a diffuse surface by modifying the kinetic energy term \cite{Lenk}.

Thomas-Fermi solutions for relevant nuclei were constructed with following force parameters.  The constants $A,B,$ and $\sigma$ (Eq.(4)) were taken to be $A$=-1533.6 MeV fm$^3$, $B$=2805.3 MeV fm$^{7/2}$, $\sigma=7/6$.  For the finite range potential we chose a Yukawa :$V_y$.
\begin{equation}
V_y=V_0\frac{e^{-|\vec{r}-\vec{r'}|/a}}{|\vec{r}-\vec{r'}|/a}
\end{equation}
with $V_0$=-668.65 MeV and $a$=0.45979 fm.  Binding energies and density profiles for many finite nuclei with these parameters (and several others) are given in Ref. \cite{Lee}. These have been used in the past to construct TF solutions which collide in heavy ion collisions \cite{Gallego}.\\

{\bf {\it Methodology:-}}
We use the method of test particles to evaluate excitation energies of a PLF with  any given shape.  The method of test particles is well-known from use of BUU models for heavy ion collisions \cite{Bertsch}.  Earlier applications were made by Wong \cite{Wong}.

We first construct a TF solution using iterative techniques \cite{Lee}.  The TF phase space distribution will then be modeled by choosing test particles with appropriate positions and momenta using Monte Carlo.  Throughout this
work we consider 100 test particles ($N_{test}=100$) for each nucleon.  For example, the phase space distribution of $^{58}$Ni is described by 5800 test particles.  A PLF can be constructed by removing a set of test particles.  Which test particles will be removed depends upon collision geometry envisaged.  For example, consider central collision of $^{58}$Ni on $^9$Be. Let $z$ to be the beam direction.  For impact parameter $b$=0 we remove all test particles in $^{58}$Ni whose distance from the center of mass of $^{58}$Ni has  $x^2+y^2<{r_9}^2$ where $r_9=2.38$ fm is the radius at half density of $^9$Be.  The cases of non-zero impact parameter can be similarly considered.

The "sudden approximation" we consider is the following.  We assume that the PLF is formed suddenly.  At the time the PLF separates from the participants the shape and momentum distribution of the PLF can be described by removing some test particles as described above.  Of course this PLF will undergo many more changes later but all we are concerned with is the energy of the system at the time of "separation".  Since the PLF now is an isolated system, the energy will be conserved.  Of course the Coulomb force from the participants will continue to be felt by the PLF.  But the major effect of this will be on overall translation of the PLF and all we are interested in is intrinsic energy.

We now describe how we calculate the energy of this "crooked" shape object.  The mass number of the PLF is the sum of the number of test particles remaining divided by $N_{test}$.  Similarly the total kinetic energy of the PLF is the sum of kinetic energies of the teat particles divided by $N_{test}$.  Evaluating potential energy requires much more work.  We need a smooth density to be generated by positions of test particles. We use the method of Lenk and Pandharipande to obtain this smooth density.  Other methods are possible \cite{Bertsch}. Experience has shown that Vlasov propagation with Lenk-Pandharipande prescription conserves energy and momenta very well \cite{Lenk} although we will not get into time propagation in this Letter.

The configuration space is divided into cubic lattices. The lattice points are $l$ fm apart.  Thus the configuration space is discretized into boxes of size $l^3$fm$^3$.  Density at lattice point $r_{\alpha}$ is defined by
\begin{equation}
\rho_L(\vec{r}_\alpha)=\sum_{i=1}^{AN_{test}}S(\vec{r}_{\alpha}-\vec{r}_i)
\end{equation}
The form factor is
\begin{equation}
S(\vec{r})=\frac{1}{N_{test}(nl)^6}g(x)g(y)g(z)
\end{equation}
where
\begin{equation}
g(q)=(nl-|q|)\Theta (nl-|q|)
\end{equation}

The advantage of this form factor is detailed in \cite{Lenk} so we will not enter that discussion here.  In this work we have always used $l$=1 fm and $n$=1.

It remains to state how we evaluate the potential energy term (eq.(4)).  The zero range Skyrme interaction contributions are very simple.  For example the first term is calculated by using
\begin{equation}
A\int d^3r\frac{\rho^2(r)}{2}=A\sum_\alpha (l^3)\rho_L^2(r_{\alpha})/2
\end{equation}
With our choice $l^3$=1 fm$^3$. For the third term in eq.(4) (the Yukawa term) is rewritten as $1/2\sum_{\alpha} \rho_L(\vec {r_{\alpha}}) \phi_L(\vec {r_{\alpha}})$ where $\phi(\vec{r})$ is the potential at $\vec{r}$ due to the Yukawa, i.e.,$\phi(\vec{r})=\int V_y(|\vec{r}-\vec{r'}|)\rho(r')d^3r'$.

The calculation of Yukawa (and/or Coulomb) potential due to a charge distribution which is specified at points of cubic lattices is very non-trivial and involves iterative procedure.  This has been used a great deal in applications involving time-dependent Hartree-Fock theory.  We will just give the references \cite{Koonin, Press, Varga}. We also found an unpublished MSU report very helpful \cite{Feldmeier}.

With this method we can calculate the total energy of the PLF.  However we are interested in excitation energy of the system which requires us to find the ground state energy of the PLF which has lost some nucleons.  We can use TF theory to find this.  The iterative TF solution also gives the ground state energy.  But then we will be using two different methods for evaluating energies, one for the PLF crooked shape and a different one for the PLF ground state. Although the results are quite close, it is more consistent to use the same prescription for ground state energy and excitation energy. Hence, even for the ground state energy we generate test particles and go through the same procedure as for the PLF with crooked shape.

However we are not finished yet.  This gives us the excitation energy but we need to know the temperature corresponding to this excitation. The canonical thermodynamic model (CTM) \cite{Das} can be used to calculate average excitation per nucleon for a given temperature, mass number and charge. This model is used to deduce the temperature for the PLF from its excitation energy. Results are shown in the next section.\\
\begin{figure} [t]
\includegraphics[width=2.0in,height=2.0in,clip]{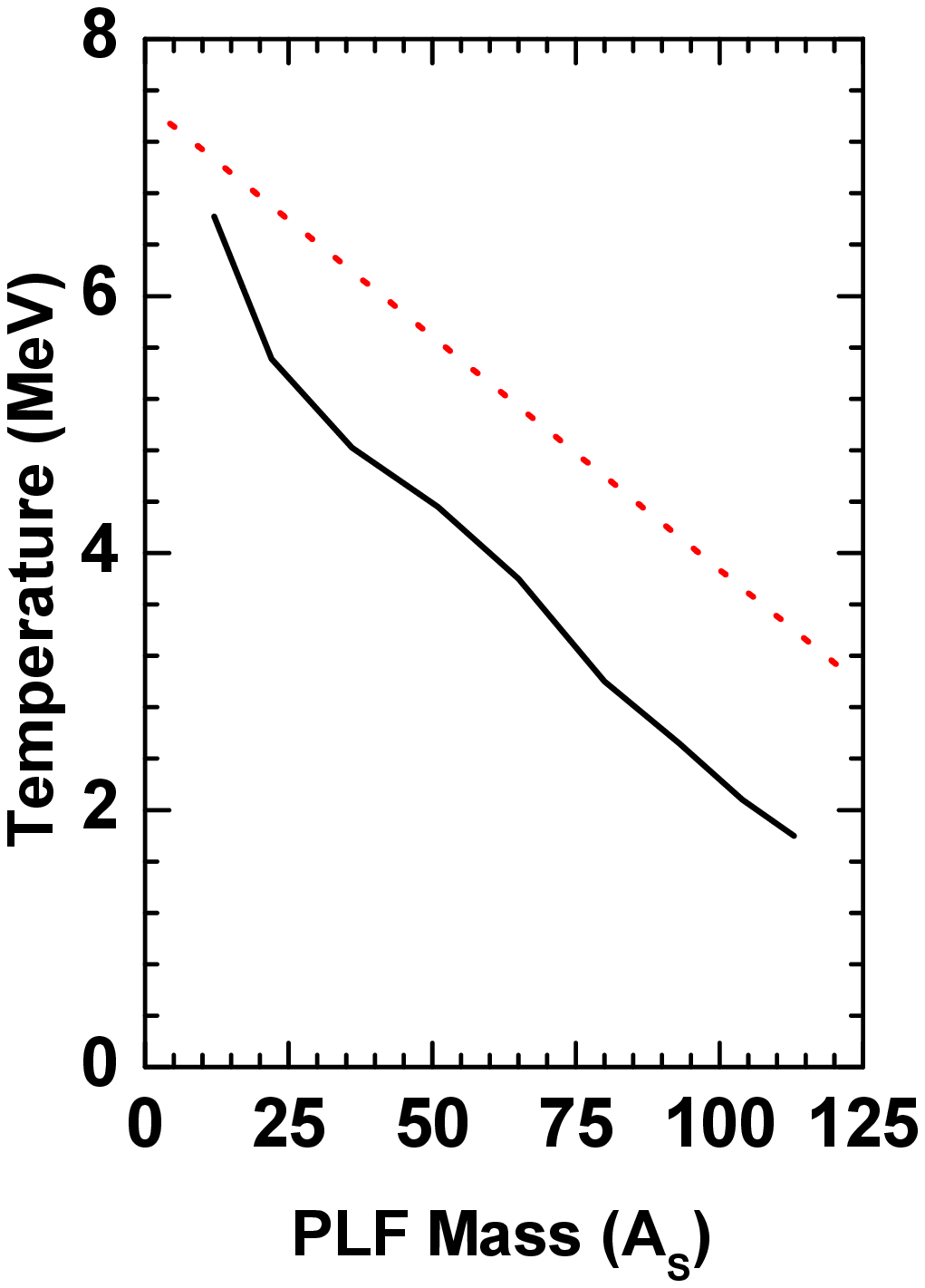}
\caption{ Plot of temperature against mass number of the PLF for $^{124}$Sn on $^{119}$Sn reaction. The solid curve is the present model, the dotted line is from our previous work \cite{Mallik3}. Since in this particular case the previous work fixed the temperature by matching with data on intermediate mass fragment production, the dotted curve represents, in some sense, "experimental" data. The dashed curve was parametrised as a function of the wound of the PLF defined as 1.0-$A_s/A_0$. The same parametrisation was used for dashed curves in Figures 3, 5 and 6.}
\end{figure}

{\bf {\it Results:-}}
\begin{figure} [t]
\includegraphics[width=3.0in,height=2.25in,clip]{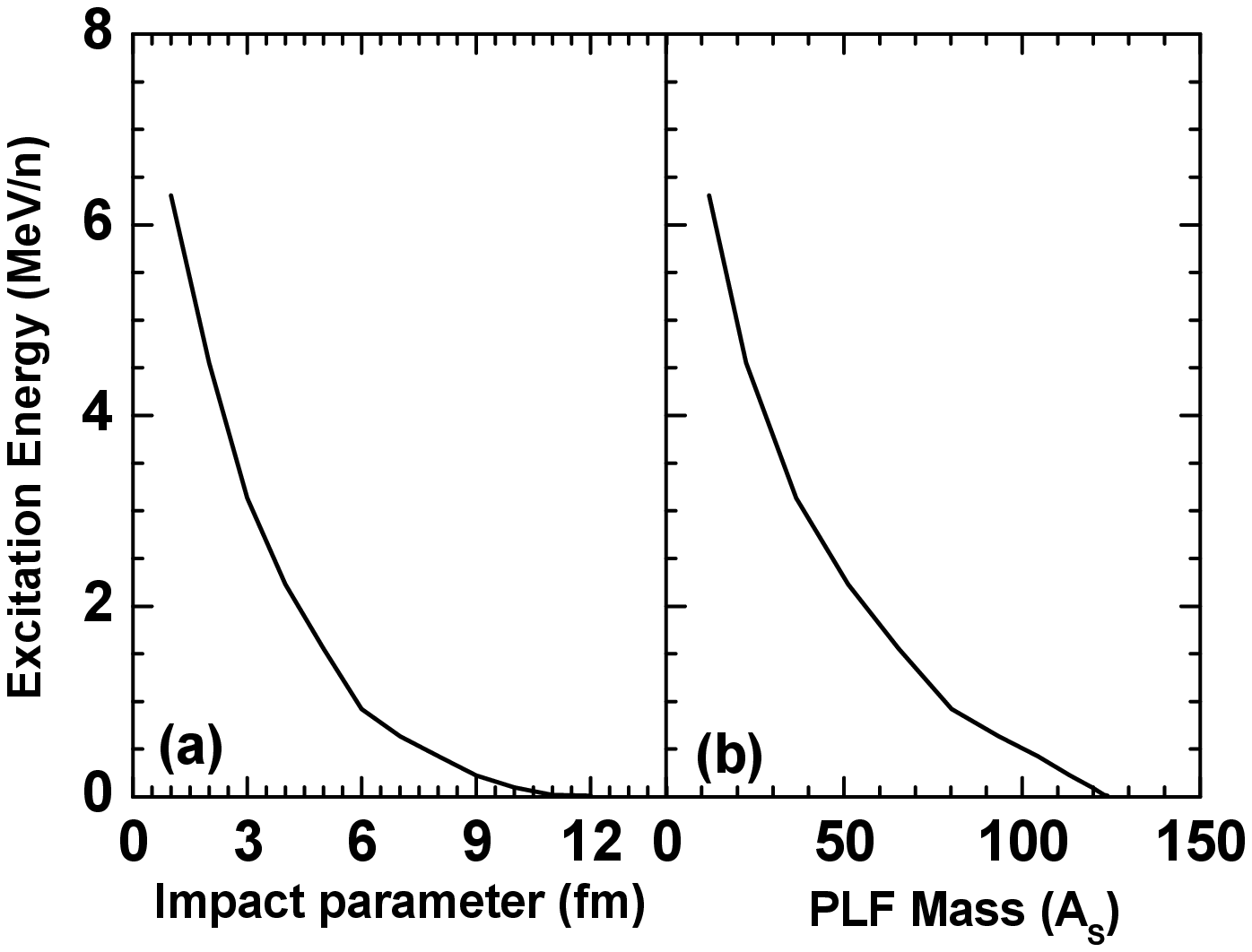}
\caption{In the model of this Letter, plot of excitation energy per nucleon for $^{124}$Sn on $^{119}$Sn against (a) impact parameter and (b) mass $A_s$ of the PLF.}
\end{figure}
We show results for $^{124}$Sn on $^{119}$Sn, $^{136}$Xe on $^{208}$Pb, $^{58}$Ni on $^9$Be and $^{181}$Ta as these were studied in our previous work. We used the data of Sn on Sn in conjunction with our model to deduce "experimental" values of temperature as a function of measured PLF mass \cite{Mallik3}.  For Sn on Sn, the total
measured charge in the forward direction specified the size of the PLF.  In the calculation we can match this size by varying the impact parameter. For this  size, experiment measured the number of intermediate mass fragments
(defined to be composites with charge between 3 and 20).  In our model, for a given size, the number of intermediate mass fragments depends on the temperature of the PLF.  We exploit this to deduce the temperature for this $b$.  Thus we are able to determine, for Sn on Sn, a temperature vs $b$ curve. However a much more useful parametrisation is in terms of the wound (=1-$A_s/A_0$) as this can be used for other colliding pairs.  We assumed a universal curve derived from the experimental curve of Sn on Sn and used that to calculate many cross-sections (not only intermediate mass
fragments) in all the cases mentioned above.  The agreements with data were very pleasing.
\begin{figure} [h]
\includegraphics[width=2.0in,height=2.0in,clip]{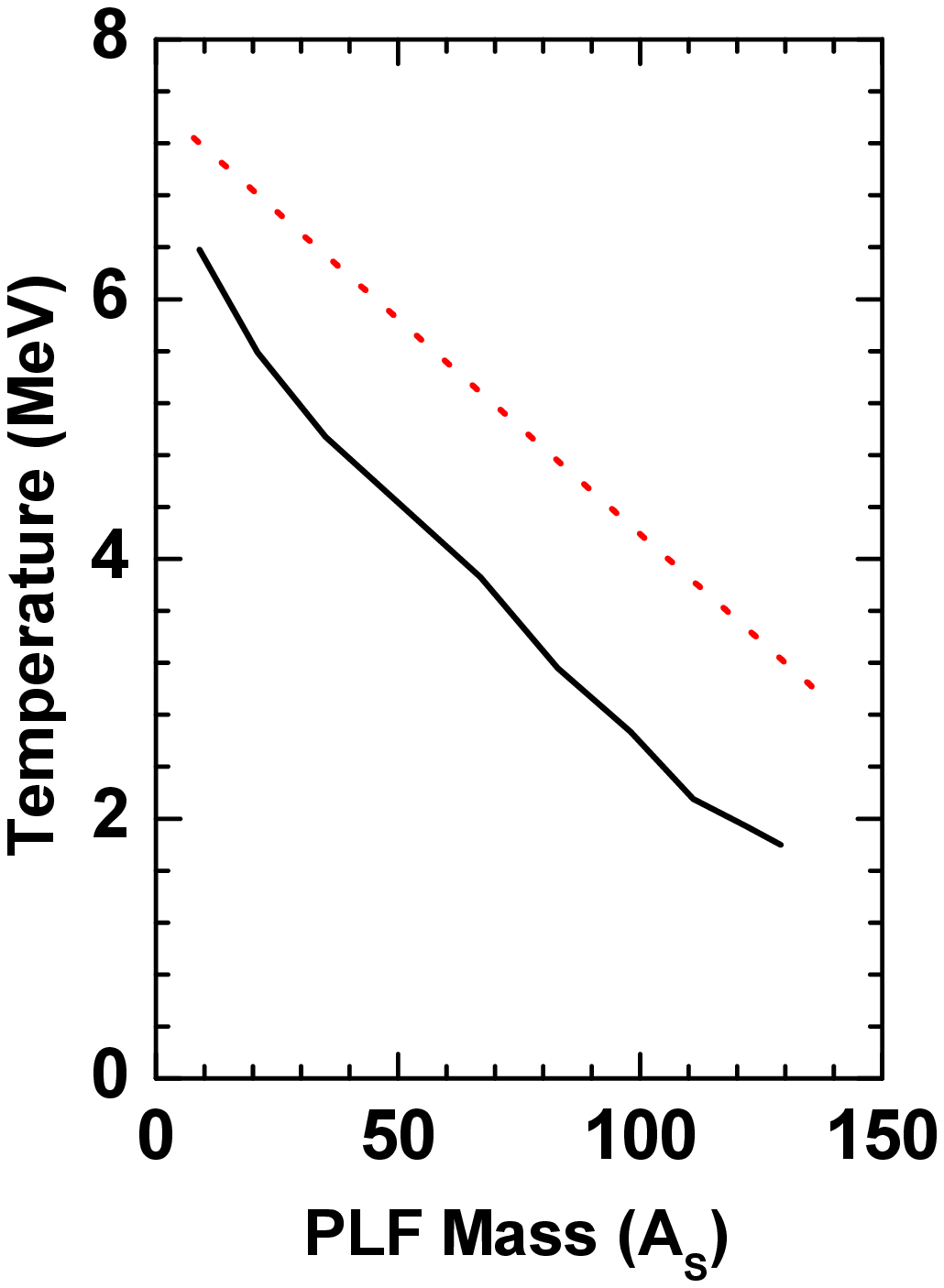}
\caption{Curve similar to that of fig.1 but for $^{136}$Xe on $^{208}$Pb.  The dotted curve is from our previous work.}
\end{figure}
In our present case, no adjustable parameters are available to deduce either the excitation energy or the temperature.  All we have is a semi-classical Hamiltonian which reproduces the ground state energy, correct radius at half density and a realistic distribution of density in finite nuclei.  It fits nuclear matter density, binding energy and compressibility. Consider the projectile $^{124}$Sn hitting the target $^{119}$Sn. We obtain the TF ground state of the projectile and create test particles with appropriate positions and momenta which map this phase-space distribution. The effect of collision with the target is that some test particles of the projectile are driven out leaving a PLF with smaller mass number and a crooked shape. A disc with the radius of the target will take out different numbers of projectile test particles for different impact parameters.  We have already described how excitation energies are calculated.

\begin{figure} [t]
\includegraphics[width=3.0in,height=2.25in,clip]{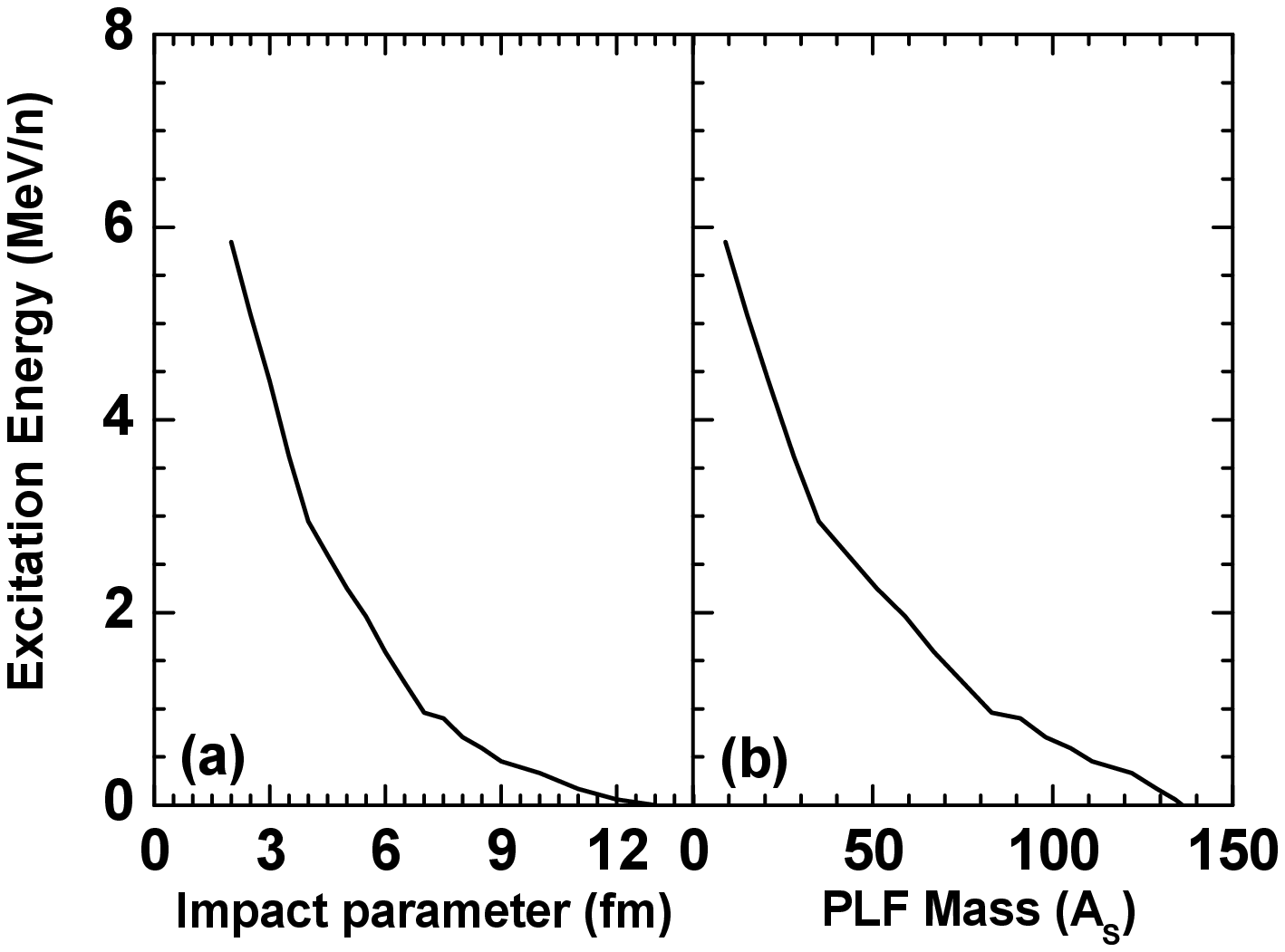}
\caption{Same as in Fig.2 except for $^{136}$Xe on $^{208}$Pb.}
\end{figure}

Figures 1 and 2 show the case of $^{124}$Sn on $^{119}$Sn.  In Figure 1 the curve showing previous calculation is essentially the experimental result as the temperatures which reproduce experimental results were selected.  Our present results obtained solely from a Hamiltonian which fits just ground state data does have the correct trend but underestimates the temperature.  This is to be expected.  We have assumed zero coupling between participants and spectators. Participants have much more energy density (hence higher temperature).  Energy flows from higher value to lower value and hence temperature of the spectator is expected to rise further.  Figure 2 plots excitation energy per nucleon against impact parameter and mass number of the PLF.  In this Letter excitation energy is directly calculated.  Temperature is deduced from excitation energy using the CTM.\\
\begin{figure} [h]
\includegraphics[width=2.0in,height=2.0in,clip]{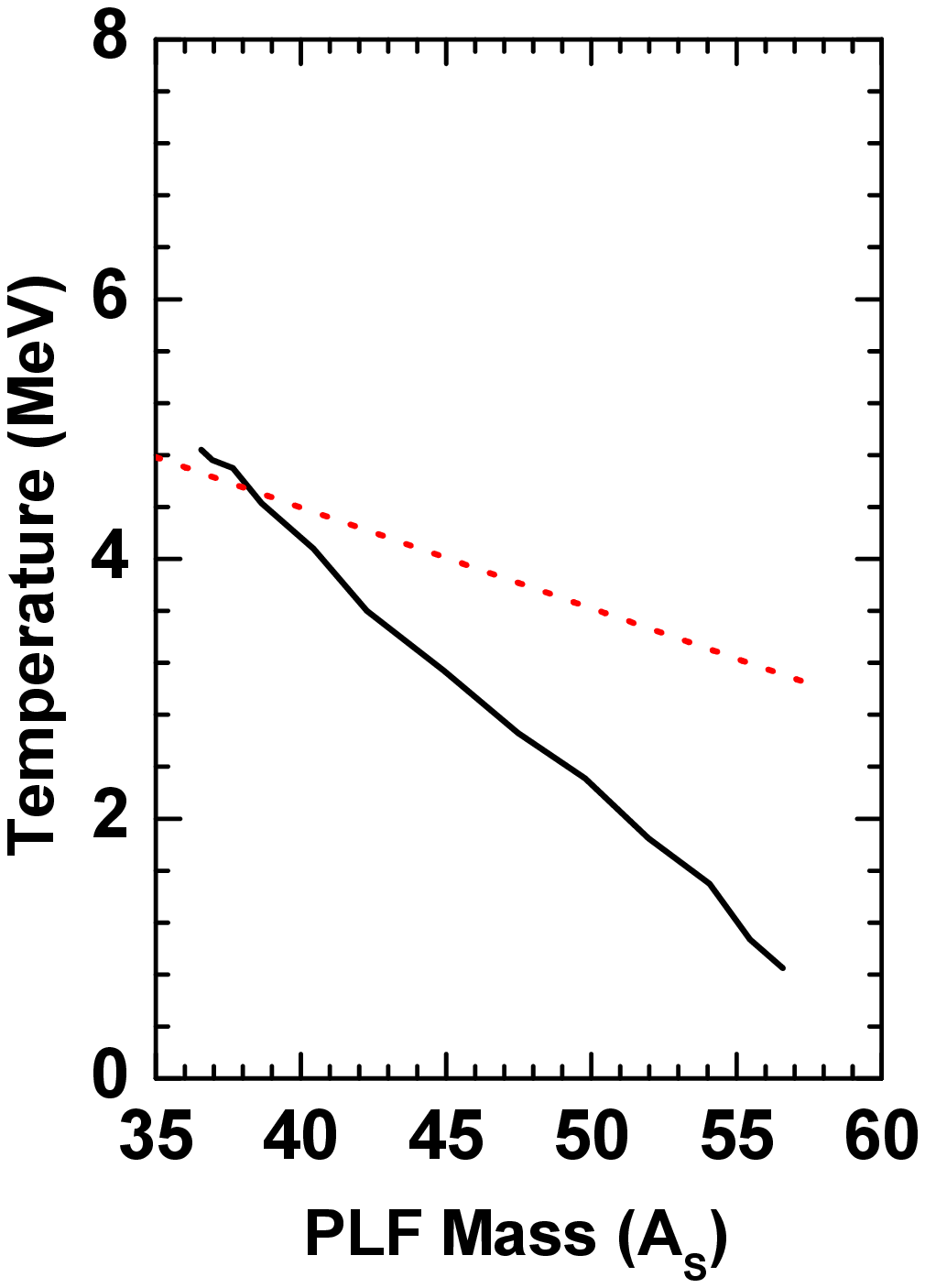}
\caption{Same as in Fig. 3 except for the case of $^{58}$Ni on $^9$Be.}
\end{figure}

\begin{figure} [t]
\includegraphics[width=2.0in,height=2.0in,clip]{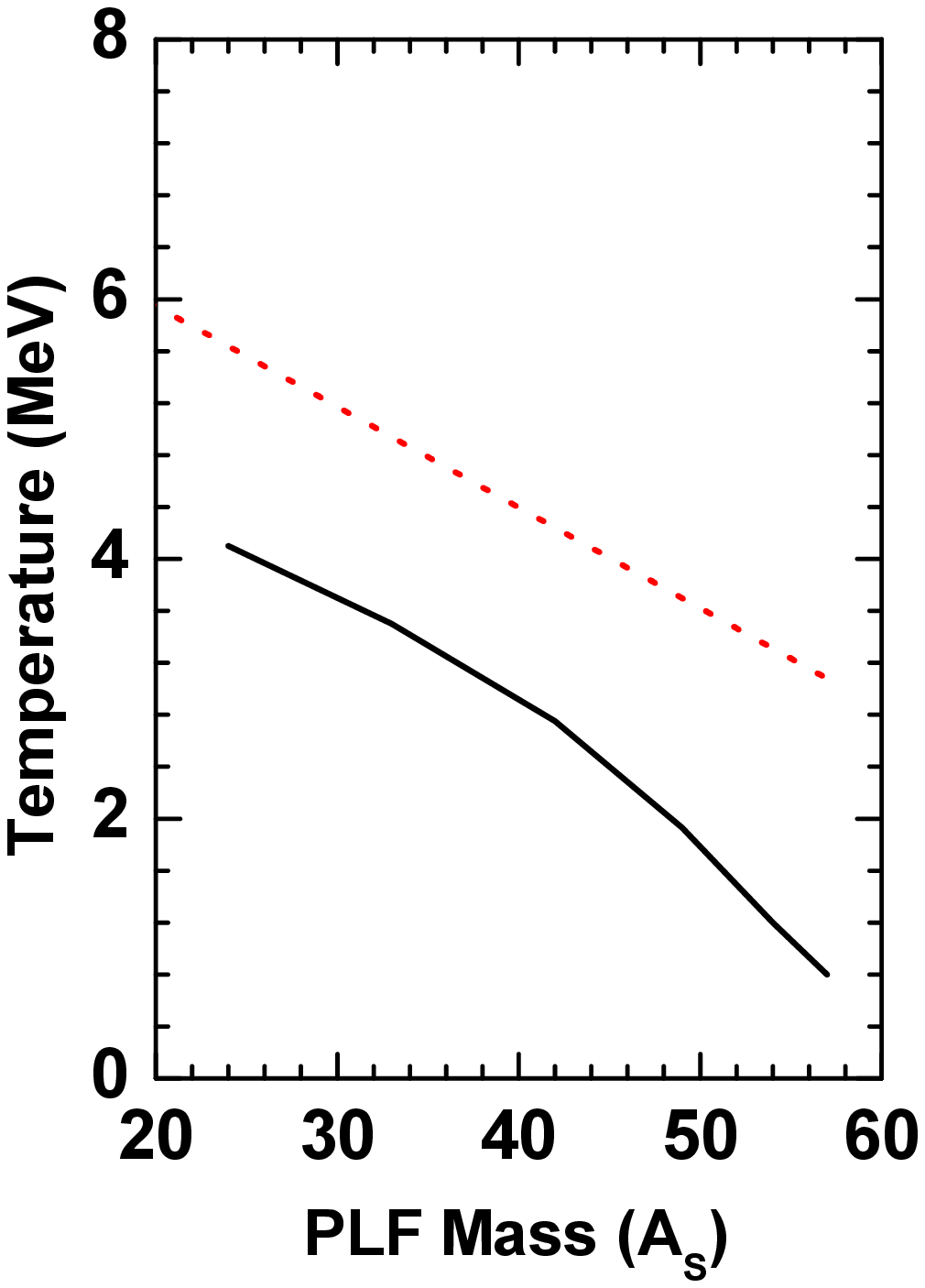}
\caption{Same as in Fig.5 except for $^{58}$Ni on $^{181}$Ta.}
\end{figure}

Figures 3 and 4 show similar results for Xe on Pb.  The functional dependence of the temperature on the "wound" is taken from Sn on Sn.  Previous calculations gave good fits with many experimental observables but they are not displayed here and can be found in ref. \cite{Mallik3, Mallik101}. Figures 5 and 6 show our results for $^{58}$Ni on $^9$Be and on $^{181}$Ta.  The results for Ni on Ta are similar to those of Sn on Sn and Xe on Pb.  For Ni on Be the present model makes the fall of temperature with growing mass $A_s$ much faster.  This may have something to do with the rather small size of the target.  For the smallest possible target, i.e., a proton, the model we have been following here is clearly inappropriate.  At what size of the target, the model becomes reasonable is hard to estimate.

Attempts to estimate excitation energy in the PLF were made earlier by Grimard and Schmidt \cite{Gaimard}. Their approach was different. It used shell model type idea. In their calculation the nucleons are bound in the potential well of the nucleus.  During abrasion, the orbits of nucleons not removed are preserved.  By abrasion, a certain number of single particle levels is vacated and the excitation energy is given by the sum of the energies of these holes with respect to the Fermi surface.  After some calculation they obtain an average excitation energy of 13.3 MeV per hole.  The goals of the two models are the same but there is no one-to-one correspondence between the two.  Arguably our model is more basic in the sense we start from a more basic interaction.  We get what semi classical mean field gives without any extra assumption.  One advantage of our approach is that the model can be seamlessly expanded to include further refinements with a BUU calculation.  Preliminary results are very encouraging.\\

{\bf {\it Acknowledgement:-}}
S. Das Gupta thanks Dr. D.K. Srivastava and Dr. A.K. Chaudhuri for hospitality during visit at Variable Energy Cyclotron Centre. He also thanks Dr. J. Gallego at McGill for many helpful discussions.

\end{document}